\documentclass[conference,english]{IEEEtran}%
\usepackage[T1]{fontenc}
\usepackage[fleqn]{amsmath}
\usepackage{graphicx}
\usepackage{amssymb}
\usepackage{amsfonts}
\usepackage{amscd}
\usepackage{multirow}
\usepackage{color}
\usepackage{babel}
\usepackage{subfigure}
\usepackage{algorithm}
\usepackage{algorithmicx}
\usepackage{listings}
\usepackage{epsfig}
\usepackage{algpseudocode}
\usepackage{array}

\providecommand{\U}[1]{\protect\rule{.1in}{.1in}}
\makeatletter

\addto\captionsenglish{}
\@ifundefined{definecolor}
{
}{}

\newtheorem{rem}{Remark}[section]

\makeatother
\begin{document}
\IEEEoverridecommandlockouts

\title{Throughput Maximization for an Energy Harvesting Communication System with Processing Cost}
\author{\IEEEauthorblockN{Oner Orhan\textsuperscript{1}\thanks{
O. Orhan and E. Erkip are supported in part by NSF CNS-0905446 and by the New York State Center for Advanced Technology in Telecommunications (CATT).}, Deniz G{\"u}nd{\"u}z\textsuperscript{2}\thanks{D. G{\"u}nd{\"u}z is supported in part by the Spanish Government under project TEC2010-17816 (JUNTOS) and the European Commission's Marie Curie IRG Fellowship with reference number 256410 under the Seventh Framework Programme.}, and Elza Erkip\textsuperscript{1}}
\IEEEauthorblockA{\textsuperscript{1}Dept. of ECE, Polytechnic Institute of New York University, Brooklyn, NY\\
\textsuperscript{2}CTTC, Barcelona, Spain\\
Email: \text{oorhan01@students.poly.edu}, \text{deniz.gunduz@cttc.es}, \text{elza@poly.edu}}
}

\maketitle

\begin{abstract}
In wireless networks, energy consumed for communication includes both the transmission and the processing energy. In this paper, point-to-point communication over a fading channel with an energy harvesting transmitter is studied considering jointly the energy costs of transmission and processing. Under the assumption of known energy arrival and fading profiles, optimal transmission policy for throughput maximization is investigated. Assuming that the transmitter has sufficient amount of data in its buffer at the beginning of the transmission period, the average throughput by a given deadline is maximized. Furthermore, a ``directional glue pouring algorithm'' that computes the optimal transmission policy is described.
\end{abstract}

\section{Introduction}
Battery size is one of the main bottlenecks on the network lifetime in wireless sensor networks. Replacing batteries may be expensive or inconvenient for nodes that are deployed in remote locations. In recent years, energy harvesting (EH) has become a viable solution to operate wireless sensor nodes in a self-powered fashion for extended periods of time. However, due to the physical and technological limitations of EH devices, harvested energy is typically low. Therefore, management of harvested energy is essential.

In wireless systems, energy consumption for communication has two components: transmission energy used by the power amplifier and the processing energy cost \cite{kim}. Depending on the range of communication or the complexity of the processing circuitry, either of these components can be the dominating factor. For an energy limited system, it is known that increasing the transmission time and lowering the transmission power as much as possible is throughput optimal when the processing energy cost is ignored \cite{verdu}. On the other hand, it is shown in \cite{glue} that when the processing energy cost is taken into account, the optimal transmission scheme becomes bursty, as increasing the transmission time means increasing the energy spent for processing.

While most of the previous work on EH communication systems focus mainly on the maximization of throughput while ignoring the processing energy cost, in this paper we consider the power used by both the power amplifier and the processing circuitry, and optimize the data transmission schedule over a fading channel. We consider a constant processing energy cost per unit time whenever the transmitter is operating. We assume that the transmitter has a finite capacity battery and that a sufficient amount of data is already available at the transmitter's data buffer before transmission starts. We focus on offline optimization, that is, the energy arrival instants and amounts as well as channel gains until the transmission deadline are known in advance by the transmitter. Note that the noncausal knowledge of energy arrivals can model accurately systems with predictable energy arrivals \cite{maria}, or provide an upper bound on the performance for the case of unpredictable energy arrivals. Our goal is to identify an optimal transmission policy that maximizes the total transmitted data by a given deadline $T$ subject to the energy causality constraint.


Recently, offline transmission policies have attracted significant interest \cite{Yang2012}-\cite{Nossek}. Optimal transmission policies which account only for the power amplifier are studied in \cite{Yang2012}-\cite{Yener}. Yang and Ulukus \cite{Yang2012} investigate a single link EH system over a constant channel for given energy and data arrival profiles, and provide an algorithm which computes the optimal transmission policy. Other communication scenarios with EH nodes that have been studied include single link fading channel \cite{fade}, multiple access channel \cite{multi}, interference channel \cite{yener2} and two-hop networks \cite{deniz}-\cite{oner}. Battery imperfections for a single link system is investigated in \cite{deniz2} and \cite{Yener}. While \cite{deniz2} studies battery leakage and reduction in battery capacity over time, \cite{Yener} investigates finite size battery. Optimal transmission policies over a constant channel while accounting for both transmission and processing energy costs are studied in \cite{process} and \cite{Nossek}.


\section{Preliminaries}\label{sec 1}
\subsection{System Model}
We consider an EH communication system that harvests energy in packets of finite amount at time instants $t_{e,0}=0<t_{e,1}<\cdots <t_{e,n}<T$ such that the packet harvested at $t=t_{e,i}$ has energy $E_i$. Harvested energy is stored in a finite battery of capacity $E_{max}$ before it is used for transmission. Therefore, without loss of generality, we can assume that each energy packet can have at most $E_{max}$ amount of energy. We assume that there is no energy loss in storing and retrieving energy from the battery. We also assume that the real valued channel gain $h(t)$ changes at time instants $t_{f,0}=0 <t_{f,1}< \cdots <t_{f,m}<T$, and remains constant in between. The channel is modeled as having additive white Gaussian noise with unit variance. Without loss of generality instantaneous transmission rate is given by Shannon capacity $\frac{1}{2}\log(1+h(t)p(t))$, where $p(t)$ is the transmission power at time $t$. We assume that the transmitter is able to change its data rate instantaneously by changing the transmission power, $p(t)$.

We can combine all energy arrivals and changes in the channel gain in a single time series $t_0=0<t_1< \cdots <t_{N-1}<T$. This can be achieved by allowing zero energy arrivals at some $t_i$'s, or the channel gain to remain constant across some of the intervals. For consistency in notation, we assume an energy arrival of $E_N=0$ at $t=T$. The time interval between two consecutive events is called an \emph{epoch}, and $\tau_i\triangleq t_{i}-t_{i-1}$ denotes the duration of the $i$'th epoch. The channel gain for epoch $i$ is denoted by $h_i$. We are interested in offline optimization, that is, we assume that the transmitter knows all the energy arrival instants and amounts as well as the channel gains for the period $0\leq t \leq T$ in advance at $t=0$.


We assume that the transmitter consumes energy only when it is transmitting, and the processing energy cost is $\epsilon$ joules per unit time, independent of the transmission power, $p(t)$. We ignore the cost of switching the transmitter `on' and `off', and assume that no information is conveyed by the state of the transmitter as in \cite{glue}.

A \emph{transmission policy} refers to a power allocation function $p(t)$ for $0\leq t\leq T$. A feasible transmission policy should satisfy the energy causality constraint:
\begin{eqnarray}
\label{const 1}
E(t) \leq \sum_{i:0\leq t_{i}<t}^{}{E_{i}},\quad \forall t\in[0,T],
\end{eqnarray}
where $E(t)$ is the total consumed energy by transmission policy $p(t)$ up to time $t$, i.e., $E(t)=\int_{0}^{t}{(p(\tau)+\epsilon\cdot \mathbf{1}_{\{p(\tau)>0\}})d\tau}$. In addition, battery overflows lead to a suboptimal transmission policy because we can always increase the throughput by increasing the transmit power such that there is no battery overflow. Therefore, an optimal transmission policy must also satisfy the following constraint:
\begin{eqnarray}
\label{const 2}
\sum_{i:0\leq t_{i}\leq t}^{}{E_{i}}- E(t) \leq E_{max}, \quad \forall t\in[0,T].
\end{eqnarray}

Assuming that the transmitter has sufficient data in its data buffer at time $t=0$, our goal is to maximize the throughput under the above constraints by deadline $T$ for given energy arrival and fading profile.

\subsection{Single Energy Arrival and Fading Level}\label{process}
It is well known that for a fixed energy budget and no processing energy, increasing the transmission duration strictly increases the throughput if the rate-power function is non-negative, strictly concave and monotonically increasing, properties that are satisfied by most of the practical rate-power functions \cite{verdu}. However, if the processing energy is not negligible, increasing the transmission duration does not increase the total amount of transmitted data after a certain point in time since the processing energy consumption starts to dominate the consumed energy. It is shown in \cite{glue} that there is an optimal transmission duration and power level which depend only on the processing energy cost $\epsilon$ and the channel gain. Note that constant power level is optimal due to the concavity of the rate-power function.

For a single energy packet arrival $E$ at time $t=0$ and a static channel state $h$, we first assume that there is no transmission deadline. Denoting the total transmission duration by $\Theta$, maximum throughput is given by the solution of the following optimization problem:
\begin{eqnarray}\label{prob 11}
\underset{\Theta, v: \Theta(v+\epsilon) \leq E}{\operatorname{max}}\; \; \frac{\Theta}{2}\log(1+h v),
\end{eqnarray}
where $v$ is the transmission power. Setting $\Theta = \frac{E}{v+\epsilon}$ in (\ref{prob 11}) and differentiating with respect to $v$, the optimal transmission power $v^*$ should satisfy
\begin{eqnarray}\label{eq 3}
\frac{1}{\frac{1}{h}+v^*}=\frac{1}{\epsilon + v^*}{\log(1+h v^*)}.
\end{eqnarray}
Above equation has only one solution for the optimal power level $v^*$ which increases as the channel gain $h$ decreases\footnote{This follows from (\ref{eq 3}) by taking the derivative of $v^*$ with respect to $h$.}. Moreover, $v^*$ does not depend on the available energy $E$. When there is a transmission deadline $T$, if $T \geq \frac{E}{v^*+\epsilon}$, then the above solution is still optimal. On the other hand, if $T<\frac{E}{v^*+\epsilon}$, transmitting at power $v^*$ cannot be optimal because some energy would remain in the battery at time $T$. In this case, we can increase the throughput by increasing the transmission power so that all the available energy is consumed by time $T$, and the optimal transmission power is given by $\frac{E}{T}-\epsilon$.

\subsection{Related Work}\label{related}
\subsubsection{Glue Pouring}\label{glue pour}
For a battery limited node with processing energy cost, when there are multiple fading levels, optimal transmission policy is different from the well-known waterfilling solution and called "glue pouring" \cite{glue}. For ease of exposure, we describe glue pouring for two fading levels, single energy arrival and no deadline. Using differential power allocation (see \cite{glue} for details), for single energy arrival $E$ and fading states $h_1>h_2$ with durations $\tau_1$, $\tau_2$, respectively, the optimal transmission policy is summarized below. In the following, $\Theta_1$ and $\Theta_2$ are transmission durations for epochs with fading levels $h_1$ and $h_2$, and $v_1^*$ and $v_2^*$ are the solutions of (\ref{eq 3}) for channel gains $h_1$ and $h_2$, respectively.
\begin{itemize}
\item If $E\leq \tau_1(v_1^*+\epsilon)$, then optimal transmission policy is $\Theta_1=\frac{E}{v_1^*+\epsilon}$ and $\Theta_2=0$ with power levels $v_1^*$ and $0$, respectively.
\item If $\tau_1(v_1^*+\epsilon)<E\leq \tau_1(v_2^*+\frac{1}{h_2}-\frac{1}{h_1}+\epsilon)$, then optimal transmission policy is $\Theta_1=\tau_1$ and $\Theta_2=0$ with power levels $\frac{E}{\tau_1}-\epsilon$ and $0$, respectively.
\item If $\tau_1(v_2^*+\frac{1}{h_2}-\frac{1}{h_1}+\epsilon)<E \leq \tau_1(v_2^*+\frac{1}{h_2}-\frac{1}{h_1}+\epsilon)+\tau_2(v_2^*+\epsilon)$, then optimal transmission policy is $\Theta_1=\tau_1$ and $\Theta_2=\frac{E-\tau_1(v_2^*+\frac{1}{h_2}-\frac{1}{h_1}+\epsilon)}{v_2^*+\epsilon}$ with power levels $v_2^*+\frac{1}{h_2}-\frac{1}{h_1}$ and $v_2^*$, respectively.
\item If $\tau_1(v_2^*+\frac{1}{h_2}-\frac{1}{h_1}+\epsilon)+\tau_2(v_2^*+\epsilon)<E$, then optimal transmission policy is the usual waterfilling.
\end{itemize}

\subsubsection{Directional Waterfilling}
\label{direc water}
The directional waterfilling algorithm, introduced in \cite{fade} for an EH fading communication system with no processing energy cost, is an adaptation of the classical waterfilling algorithm to the EH model where the energy becomes available over time. Due to energy causality harvested energy $E_i$ can only be allocated to epochs $j> i$; and, due to the battery constraint, the amount of energy that can be transferred to epoch $j$ is limited by $E_{max}-E_{j-1}$.

\section{Throughput Maximization}
\label{sec 3}
\subsection{Problem Formulation and Solution}
\label{problem form}
In this section we study the throughput maximization problem with multiple energy arrivals and fading levels.
It is possible to show that within each epoch, when the transmitter is `on', constant power transmission is optimal \cite{Yang2012}, so we denote the nonnegative power level within epoch $i$ as $p_i$ with duration $\Theta_i$, $0 \leq \Theta_i \leq \tau_i$. Then, the throughput optimization problem can be stated as follows:
\begin{subequations}
\label{prob 1}
\begin{align}\label{prob 1:11}
&\underset{p_i,\Theta_i}{\operatorname{max}} && \sum_{i=1}^{N}{\frac{\Theta_i}{2} \log(1+h_ip_i)} \\\label{prob 1:1}
&\text{s.t.} && 0\leq \sum_{j=1}^{i}{(E_{j-1}-\Theta_j (p_j+\epsilon))},  i=1,...,N, \\\label{prob 1:4}
&&&\sum_{j=1}^{i+1}{E_{j-1}}-\sum_{j=1}^{i}{\Theta_j (p_j+\epsilon)}\leq E_{max},   i=1,...,N, \\ \label{prob 1:2}
&&&0\leq \Theta_i \leq \tau_i, ~~\mbox{ and } ~~0\leq p_i,~~i=1,...,N.
\end{align}
\end{subequations}
Note that this is not a convex optimization problem because the constraints in (\ref{prob 1:1})-(\ref{prob 1:4}) are not convex. Therefore, we will reformulate this problem by defining a new variable $\alpha_i \triangleq \Theta_i p_i$, which denotes the total consumed energy by the power amplifier within epoch $i$. Then, the optimization problem in (\ref{prob 1}) can be written in terms of $\Theta_i$ and $\alpha_i$ as follows:
\begin{subequations}\label{prob 3}
\begin{align}\label{prob 3:11}
&\underset{\alpha_i,\Theta_i}{\operatorname{max}} && \sum_{i=1}^{N}{\frac{\Theta_i}{2} \log\left(1+\frac{h_i \alpha_i}{\Theta_i}\right)} \\\label{prob 3:1}
&\text{s.t.} && 0\leq \sum_{j=1}^{i}{(E_{j-1}-\alpha_j-\epsilon \Theta_j)},  \quad i=1,...,N, \\\label{prob 3:2}
&&&\sum_{j=1}^{i+1}{E_{j-1}}-\sum_{j=1}^{i}{(\alpha_j+\epsilon \Theta_j)}\leq E_{max}, i=1,...,N, \\\label{prob 3:3}
&&&0\leq \Theta_i \leq \tau_i, ~~\mbox{ and } ~~ 0\leq \alpha_i,  ~~i=1,...,N.
\end{align}
\end{subequations}

With this reformulation, concavity of (\ref{prob 3:11}) can be argued from the fact that the function $\frac{\Theta_i}{2} \log(1+\frac{h_i \alpha_i}{\Theta_i})$ is the perspective of the strictly concave function $\frac{1}{2} \log(1+h_i \alpha_i)$. Since perspective operation preserves concavity, $\frac{\Theta_i}{2} \log(1+\frac{h_i \alpha_i}{\Theta_i})$ is also concave \cite{Boyd}. The linear constraints in (\ref{prob 3:1})-(\ref{prob 3:3}) define a convex feasible set, therefore, the optimization problem in (\ref{prob 3}) is a convex optimization problem.

The Lagrangian of (\ref{prob 3}) with $\lambda_i \geq 0$, $\mu_i \geq 0$, $\gamma_i \geq 0$, $\nu_i \geq 0$ and $\sigma_i \geq 0$ can be written as:
\begin{align}\label{lagran 1}
\mathcal{L} & = \sum_{i=1}^{N}{\frac{\Theta_i}{2} \log\left(1+\frac{h_i \alpha_i}{\Theta_i}\right)}\\
               &-\sum_{i=1}^{N}{\lambda_i \left(\sum_{j=1}^{i}{(\alpha_j+\epsilon \Theta_j-E_{j-1})}\right)}\nonumber\\
               & -\sum_{i=1}^{N}{\mu_i \left(\sum_{j=1}^{i+1}{E_{j-1}}-\sum_{j=1}^{i}{(\alpha_j+\epsilon \Theta_j)}-E_{max}\right)} \nonumber \\
               & -\sum_{i=1}^{N}{\gamma_i (\Theta_i-\tau_i)} +\sum_{i=1}^{N}{\nu_i \Theta_i} +\sum_{i=1}^{N}{\sigma_i \alpha_i}.\nonumber
\end{align}
Corresponding complementary slackness conditions are
\begin{align}\label{comp 1:1}
{\lambda_i \left(\sum_{j=1}^{i}{(\alpha_j+\epsilon \Theta_j-E_{j-1})}\right)}&= 0, ~ \forall i\\\label{comp 1:2}
{\mu_i \left(\sum_{j=1}^{i+1}{E_{j-1}}-\sum_{j=1}^{i}{(\alpha_j+\epsilon \Theta_j)}-E_{max}\right)}&= 0, ~ \forall i\\\label{comp 1:3}
{\gamma_i (\Theta_i-\tau_i)} = 0, ~~ {\nu_i \Theta_i} = 0 ~~ \mbox{ and }
~~{\sigma_i \alpha_i}=&0, ~~ \forall i.
\end{align}
Taking derivatives with respect to $\alpha_i$ and $\Theta_i$, we obtain
\begin{eqnarray}
\label{der 1}
\frac{\partial \mathcal{L}}{\partial \alpha_i} = \frac{\Theta_i h_i}{2(\Theta_i +h_i \alpha_i)}- \sum_{j=i}^{N}{(\lambda_j-\mu_j)} + \sigma_i,
\end{eqnarray}
\begin{eqnarray}
\label{der 2}
\frac{\partial \mathcal{L}}{\partial \Theta_i} &=& \frac{1}{2}\log\left(1+\frac{h_i \alpha_i}{\Theta_i}\right)- \frac{h_i \alpha_i}{2(\Theta_i + h_i \alpha_i)} \nonumber\\
&&-\epsilon \sum_{j=i}^{N}{(\lambda_j-\mu_j)}-\gamma_i+ \nu_i.
\end{eqnarray}
We consider KKT conditions together with the complementary slackness conditions in (\ref{comp 1:1})-(\ref{comp 1:3}).
\begin{itemize}
\item If $\Theta_i^*=0$, then $\alpha_i^*=0$ and no power is allocated to epoch $i$, i.e., $p_i^*=0$.
\item If $0<\Theta_i^* < \tau_i$ and $0 < \alpha_i^*$, i.e., $\gamma_i = 0$, $\nu_i =0$, $\sigma_i=0$, then from (\ref{der 1}) and (\ref{der 2}), we obtain
\begin{eqnarray}
\label{eq 4}
\log\left(1+\frac{h_i \alpha_i^*}{\Theta_i^*}\right)= \frac{h_i(\alpha_i^* + \epsilon \Theta_i^*)}{\Theta_i^* +h_i \alpha_i^*},
\end{eqnarray}
which is equivalent to (\ref{eq 3}) when we replace $\alpha_i^*$ with $\Theta_i^* p_i$. Therefore, as argued in Section \ref{process}, (\ref{eq 4}) has a unique solution which depends only on $h_i$ and $\epsilon$. We denote this unique power level as $p_i^*=v_i^*$ and note that $p_i^*$ does not depend on the Lagrange multipliers $\lambda_j$ and $\mu_j$, for $j=i,...,N$, either.
\item If $\Theta_i^* = \tau_i$ and $0 < \alpha_i^*$, i.e., $\gamma_i > 0$, $\nu_i =0$, $\sigma_i=0$, then from (\ref{der 1}), we get
\begin{eqnarray}
\label{eq 5}
\frac{\Theta_i^* h_i}{2(\Theta_i^* +h_i \alpha_i^*)}=\sum_{j=i}^{N}{(\lambda_j-\mu_j)}.
\end{eqnarray}
Similarly, from (\ref{der 2}), we get
\begin{eqnarray}
\label{eq 6}
\frac{1}{2}\log\left(1+\frac{h_i \alpha_i^*}{\Theta_i^*}\right)=\frac{h_i \alpha_i^*}{2(\Theta_i^* + h_i \alpha_i^*)}\nonumber \\
+\epsilon \sum_{j=i}^{N}{(\lambda_j-\mu_j)}+\gamma_i.
\end{eqnarray}
Since $\gamma_i>0$, from (\ref{eq 5}) and (\ref{eq 6}), we can obtain the following inequality:
\begin{eqnarray}
\label{eq 7}
\log\left(1+\frac{h_i \alpha_i^*}{\Theta_i^*}\right) > \frac{h_i(\alpha_i^* + \epsilon \Theta_i^*)}{\Theta_i^* +h_i \alpha_i^*}.
\end{eqnarray}
Comparing with (\ref{eq 4}) we conclude that (\ref{eq 7}) is satisfied only if $p_i^*>v_i^*$. Then, by using (\ref{eq 5}) and replacing $\alpha_i^*$ with $\Theta_i^* p_i^*$, we can compute the optimal power level $p_i^*$ as
\begin{eqnarray}
\label{eq 8}
 p_i^* =\frac{1}{2\sum_{j=i}^{N}{(\lambda_j-\mu_j)}}-\frac{1}{h_i} \quad \text{if } p_i^*>v_i^*.
\end{eqnarray}
Notice that $\lambda_j$ and $\mu_j$ cannot be positive simultaneously. Using the complementary slackness conditions in (\ref{comp 1:1}) and (\ref{comp 1:2}), we have $\lambda_i>0$ and $\mu_i=0$ whenever the battery of the transmitter depletes. Therefore, from (\ref{eq 8}), we can argue that $p_{i+1}^*+\frac{1}{h_{i+1}}>p_{i}^*+\frac{1}{h_{i}}$. This means that whenever the sum of the inverse channel gain and the optimal power level increases from one epoch to the next, the battery must be empty. When $\mu_i>0$ and $\lambda_i=0$, the battery is full. Therefore, from (\ref{eq 8}), we can argue that $p_{i+1}^*+\frac{1}{h_{i+1}}<p_{i}^*+\frac{1}{h_{i}}$. As a result, the sum of the inverse channel gain and the optimal power level decreases from one epoch to the next, whenever the battery is full. Moreover, since depleting all the harvested energy by the deadline is optimal \cite{Yener}, $\lambda_N>0$ and $\mu_N=0$.
\end{itemize}
\begin{rem}
\label{remark 2}
Optimization problem in (\ref{prob 3}) may have multiple solutions. Consider a channel with multiple epochs having the same channel gain $h_i$ and a corresponding optimal transmission policy with $0<\Theta_i^* < \tau_i$. As argued from KKT conditions, $p_i^*=v_i^*$ must be satisfied for these epochs. Then the corresponding optimal values for $\frac{h_i \alpha_i^*}{\Theta_i^*}=h_i p_i^*$ must also be the same. Without loss of optimality, we can find another optimal transmission policy by transferring some of the energy between those epochs such that the optimal $h_i p_i^*$ is preserved. Note that if the channel gains are different (Section \ref{process}) or $p_i^*>v_i^*$, i.e., $\Theta_i^*=\tau_i$ or $\lambda_i>0$ or $\mu_i>0$, for all $i$, then there is a unique solution.
\end{rem}

\subsection{Directional Backward Glue Pouring Algorithm}
\label{algot1}
We can allocate the harvested energy to epochs starting from the last non-zero energy packet to the first such that constraints in (\ref{const 1})-(\ref{const 2}) (akin to directional waterfilling in Section \ref{related}) are satisfied. In addition, the optimal transmission policy utilizes epoch $i$ either partially, i.e. $\Theta_i^*<\tau_i$, with power level $p_i^*=v_i^*$, or fully, i.e., $\Theta_i^*=\tau_i$, with power level $p_i^*>v_i^*$ as argued in Section \ref{problem form}.
Therefore, the optimal transmission policy is a directional glue pouring algorithm in which each harvested energy packet $E_i$ is allocated to subsequent epochs using the glue pouring algorithm.


\begin{figure}[!ht]
\centering
\subfigure[]{
\includegraphics[scale=0.45,trim= 0 15 0 4]{pic1.eps}
\label{fig 4:subfig1}
}
\subfigure[]{
\includegraphics[scale=0.45,trim= 0 15 0 4]{pic2.eps}
\label{fig 4:subfig2}
}
\subfigure[]{
\includegraphics[scale=0.45,trim= 0 15 0 4]{pic3.eps}
\label{fig 4:subfig3}
}
\caption{Directional backward glue pouring algorithm}
\label{fig 4}%
\end{figure}

Consider the example in Fig. \ref{fig 4:subfig1}. Arrival times of the harvested energy packets are shown with thick downward arrows. The thin downward arrows correspond to the time instants when the channel gain changes with virtual energy arrivals ($E_1=E_3=E_4=0$). Inverse of the channel gains are indicated by solid blocks. Optimal power levels $v_i^*$ for $0<\Theta_i<\tau_i$ are indicated with dashed horizontal lines above the inverse channel gain blocks such that $v_i^*$ corresponds to the distance between the dashed lines and solid blocks. Note that the energy consumed for processing is not shown in the figure; however, it can be computed from the total transmission duration. As argued before, the algorithm first computes the optimal power level for the last non-zero energy arrival $E_2$. As shown in Fig. \ref{fig 4:subfig2}, the algorithm considers the harvested energy $E_2$ for epochs three, four and five. It allocates $E_2$ to the third and fourth epochs using glue pouring algorithm as argued in Section \ref{related}. Note that only a portion of the third epoch is utilized with power level $v_3^*$ due to glue pouring. Then, the algorithm considers the first non-zero energy arrival $E_0$ and allocates this energy according to glue pouring algorithm as shown in Fig. \ref{fig 4:subfig3}. Note that some of the energy is transferred to third and fourth epochs as argued in Section \ref{related}. However, transferred energy is limited due to the finite battery size which explains the water level difference (Section \ref{sec 3}).
\begin{figure}[t]
\centering
\subfigure[]{
\includegraphics[scale=0.65,trim= 10 0 0 0]{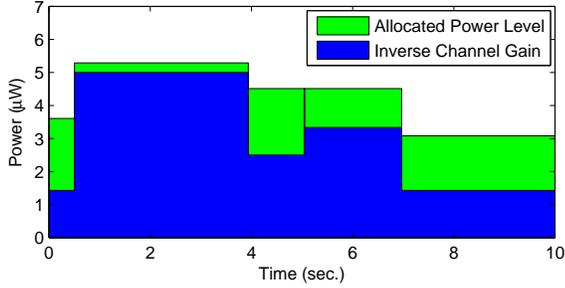}
\label{fig 7:subfig1}
}
\subfigure[]{
\includegraphics[scale=0.65,trim= 10 0 0 0]{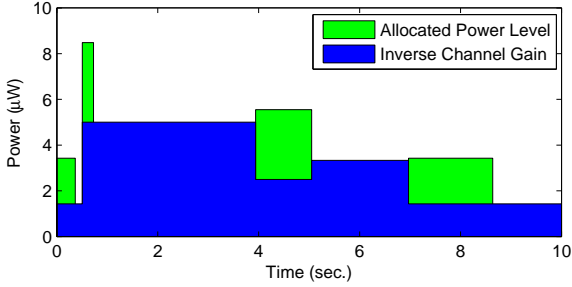}
\label{fig 7:subfig3}
}
\caption{(a) Optimal power levels for $\epsilon=0$ are $\mathbf{P}=[2.17,0.29,2.01,$ $1.18,1.65]\; \mu W$ with durations $\mathbf{\Theta}=[0.5,3.5,1.1,1.9,3.0]$ sec. Total transmitted data is $B=2.11$ nats. (b) Optimal power levels for  $\epsilon=1\; \mu W$ are $\mathbf{P}=[1.99,3.48,3.05,0,1.99]\; \mu W$ with durations $\mathbf{\Theta}=[0.36,0.22,1.10,0,1.66]$ sec. Total transmitted data is $B=1.39$ nats.}
\label{fig 7}%
\vspace{-0.1in}
\end{figure}

\section{Numerical Results}
\label{numerical}
In this section, we provide numerical results to show the effect of processing energy cost on the optimal throughput. We consider an energy arrival profile $\mathbf{E}=[1.1,3.2,2.8,1.4, 3.1]\;$ microjoules ($\mu J$), channel gains $\mathbf{h}=[0.7,0.2,0.4,0.3,0.7]\times 10^6$ with epoch durations $\mathbf{\tau}=[0.5,3.5,1.1,1.9,3.0]$ sec until deadline $T=10$ sec. We consider that energy packets arrive at time instants when channel gain changes. We set $E_{max}=5\; \mu J$. The optimal offline transmission policy with no processing energy cost, i.e., $\epsilon=0$ is shown in Fig. \ref{fig 7:subfig1}. As we can see from Fig. \ref{fig 7:subfig1}, the transmitter utilizes each epoch fully. The difference in power levels is due to the energy causality and finite battery capacity. Using the same energy arrival and channel profile for $\epsilon=1\; \mu W$, we obtain the transmission policy in Fig. \ref{fig 7:subfig3}. As shown in the figure, the optimal transmission policy is bursty, and the total transmission energy is reduced due to the processing energy cost. Notice that the optimal policy allocates energy to epoch two even though it has the worst channel gain. This is due to finite capacity battery.

\begin{figure}[t]
\centering
\includegraphics[scale=0.63,trim= 0 0 0 0]{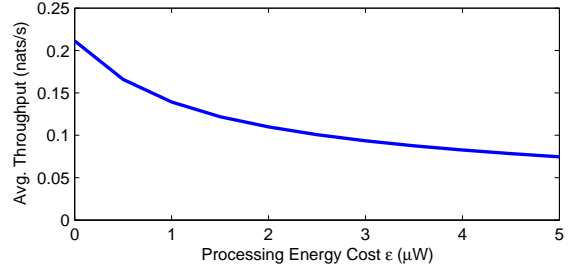}
\caption{Average throughput versus processing energy cost.}
\label{fig 8}%
\vspace{-0.1in}
\end{figure}

The variation of the throughput with respect to $\epsilon$ for the same energy and channel profile given above is shown in Fig. \ref{fig 8}. As expected the optimal throughput decreases as the processing energy cost increases.

\section{Conclusions}
\label{conclude}
In this paper, we have studied an EH communication system with processing energy cost over a fading channel. Under the noncasual knowledge of energy packet arrivals, we have identified the optimal transmission policy such that the transmitted data is maximized by a given deadline. Our solution involves a convex optimization formulation of the problem as well as an optimal `directional glue pouring' algorithm. Finally, numerical results have been provided to illustrate the effect of processing energy cost on the throughput.

\end{document}